\documentstyle[10pt]{article}

\def\bfl{\begin{flushleft}}
\def\efl{\end{flushleft}}
\def\bfr{\begin{flushright}}
\def\efr{\end{flushright}}
\def\bc{\begin{center}}
\def\ec{\end{center}}
\def\be{\begin{equation}}
\def\ee{\end{equation}}
\def\ba{\begin{eqnarray}}
\def\ea{\end{eqnarray}}

\def\lb#1{\label{#1}}

\def\text#1{\mbox{#1}}
\def\drm{\text{d}}

\def\Sign#1{\, \text{sgn}\left[#1\right] }
\def\PDer#1#2{\,\frac{\partial #1}{\partial #2}}

\begin{document}

~\\
\bfl
{\it General Relativity and Gravitation, Vol. 31, No. 4, 
pp. 571-577, 1999}
~\\
~\\
~\\
~\\
~\\
~\\
~\\
~\\
~\\
~\\
{\LARGE \bf LETTER}
~\\
~\\
\efl
~\\
\bfl
{\LARGE \bf
Classical and Quantum Evolution of Non-Isentropic

Hot Singular Layers in Finite-Temperature \vspace{2mm}

General Relativity
}

~\\
~~\\
{\large \bf
Konstantin G. Zloshchastiev\footnote{Postal address: Box 2837,
Dnepropetrovsk 320128,
Ukraine. E-mail: zlosh@email.com.
URL(s): http://zloshchastiev.webjump.com, http://zloshchastiev.cjb.net}}\\

\efl

\bc
{\it Received December 14, 1998}
\ec

\abstract{\normalsize
The spherically symmetric layer of matter
is considered within the frameworks of general relativity.
We perform generalization of the already known theory for the 
case of nonconstant surface entropy and finite temperature.
We also propose the minisuperspace model to determine the behavior
of temperature field and perform the Wheeler-DeWitt quantization.
}

~\\

PACS number(s): 04.20.Fy, 04.40.Nr, 04.60.Kz, 11.27.+d\\


~\\

\large

Beginning with the outstanding classic works 
(see Ref. \cite{isr} and references therein)
the theory of singular hypersurfaces
has been intensively developed both in the axiomatic and applied 
aspects \cite{mtw}.
Regrettably, the majority of recent investigations (especially quantum) 
mainly touch upon the subject of
the temperatureless theory of isentropic thin shells.
However, considering thermal effects  should always
increase the physical relevance 
of any approach and expand the domain of its applicability. 
Therefore, 
the aim of present letter is to study the non-isentropic singular
layers at finite temperature within the frameworks of 
classical and quantum general relativity.

So, one considers an infinitely thin layer of matter with the 
thermally generalized surface stress-energy 
tensor of a perfect fluid 
\be                                                          \lb{eq1}
{\cal S}_{ab}=(\sigma - \frac{1}{A} \int T \drm S) u_a u_b 
+ p (u_a u_b +~ ^{(3)}\!g_{ab}),
\ee
where $\sigma$ and $p$ are respectively the surface energy density and 
pressure, $u^a$ is the timelike unit tangent vector, 
$^{(3)}\!g_{ab}$ is the three-dimensional world sheet metric on the layer;
$A$, $S$ and $T$ are surface area, entropy and temperature respectively. 

Assuming the layer $\Sigma$ to be spherical 
we suppose the metrics of spacetimes 
outside $\Sigma^+$ and inside $\Sigma^-$ in the special form
\be
\drm s_\pm^2 =
-\Phi^{\pm}(r) \drm t^2_\pm + \Phi^{\pm}(r)^{-1} \drm 
r^2 + r^2 \drm \Omega^2,                                      \label{eq2}
\ee
where $d\Omega^2=\drm\theta^2+\text{sin}^2\theta \drm \varphi^2$.
Of course, with this we have some loss of generality but the assumption
(\ref{eq2}) turns to be enough for major physically interesting tasks.
In terms of the layer's proper time $\tau$ the world sheet three-metric 
is
\be
^{(3)}\!\drm s^2 = - \drm \tau^2 + R^2 \drm \Omega^2,        \label{eq3}
\ee
where $R=R(\tau)$ is a proper radius of the layer.
The Einstein equations can be decomposed as the jump of extrinsic 
curvatures across the singular layer
\be
(K^a_b)^+ - (K^a_b)^- = 4 \pi\sigma (2 u^a u_b + \delta^a_b).  \lb{eq4}
\ee
For the spacetimes (\ref{eq2}) after straightforward computing
the $\theta\theta$ component of this equation yields the equation
of layer's radial motion
\be                                                          \lb{eq5}  
\epsilon_+ \sqrt{\dot R^2 + \Phi^+}-
\epsilon_- \sqrt{\dot R^2 + \Phi^-} =- m/R,
\ee
where $\Phi^\pm = \Phi^\pm (r)|_{r=R(\tau)}$, 
$\dot R=\drm R/\drm\tau$ is the radial velocity, 
$\epsilon_\pm = \Sign{\sqrt{1+\dot R^2+\Phi^\pm}}$, 
$m=4\pi\sigma R^2$ is interpreted as the (effective) rest mass.
The sign $\epsilon = +1$ if $R$ increases in 
the outward normal of a layer, and $\epsilon = -1$ if $R$ decreases.
Besides, double squaring we can write eq. (\ref{eq5}) in the form
\be
\dot R^2 = 
\left[
\frac{\Delta \Phi - m^2/R^2}
     {2 m/R}
\right]^2 - \Phi^-,                                          \label{eq6}
\ee
where $\Delta \Phi = \Phi^+ - \Phi^-$.

The necessary condition of integrability of the Einstein equations is
the energy conservation law ${\cal S}^a_{b;a}=0$ 
for matter on the layer.
Using eqs. (\ref{eq1}), (\ref{eq3}) and the property
\be                                                           \lb{eq7}
\frac{4\pi}{\text{sin}\theta}  \sqrt{|\det{(~^{(3)}\!g_{ab})}|}  =
4\pi R^2 = A,
\ee
it can be represented as the first thermodynamical law
\be
\drm \left( \sigma A \right) +
p~ \drm A - T \drm S
+ A\,  \Delta T^{\tau n}\, \drm \tau =0,               \label{eq8}
\ee
where $\Delta T^{\tau n} = (T^{\tau n})^+ - (T^{\tau n})^-$,  
$T^{\tau n}=T^{\alpha\beta} u_\alpha n_\beta$ is the
projection of the stress-energy tensors in the $\Sigma^\pm$
spacetimes on the tangent and normal vectors.
It can be checked immediately that for spacetimes (\ref{eq2})
$T^{\tau n} \equiv 0$, and we have a conservative system
\be
\drm \left( \sigma A \right) +
p~ \drm A - T \drm S =0.               \label{eq9}
\ee
We will assume the layer's 
temperature as the internal degree of freedom which appears
to be complementary to radius (or, equivalently, to area).
Following the definition of the entropy as a state function of 
temperature and area, we obtain
\be                                                        \lb{eq10}
\drm S = \PDer{S}{T} \drm T + \PDer{S}{A} \drm A.
\ee
Comparing eqs. (\ref{eq9}) and (\ref{eq10}), we obtain
\ba
&&\PDer{S}{T} = \frac{A}{T} \PDer{\sigma}{T},                  \lb{eq11}\\
&&\PDer{S}{A} = \frac{1}{T} 
             \left[
                   p + \PDer{(\sigma A)}{A}
             \right].                                        \lb{eq12}
\ea
Then the equality of mixed derivatives yields the partial
differential equation
\be                                                          \lb{eq13}
\PDer{\sigma}{A} - T \PDer{p}{T} + p + \sigma = 0,
\ee
which is useful both for obtaining the internal energy as a function
of area and temperature from a known equation of state and for the 
inverse problem.
Thus, eqs. (\ref{eq5}), (\ref{eq9}) and (\ref{eq13}) together with an
equation of state and choice of signs $\epsilon^\pm$ are 
{\it almost} sufficient for finding of all the unknowns.
The equation completing this system 
is that for temperature field,
and seems to be introduced from some additional assumptions.
One of them is the variational minisuperspace model which will
be proposed below.

Let us choose now an equation of state of layer's 2D matter.
The simplest (but physically most interesting) EOS is the linear one
of barotropic fluid,
\be                                                          \lb{eq14}
p = \eta \sigma,
\ee
then the quasilinear PDE (\ref{eq13}) has the general solution
\be                                                          \lb{eq15}
\sigma = f (A^\eta T) A^{-\eta-1},
\ee
where $f$ is an arbitrary function which can be determined by means of 
initial conditions, correspondence principle, etc.
For instance, at $\eta \not= 0$ one can assume
\be                                                          \lb{eq16}
f^{(ph)} (x) = \alpha + \beta x^{1+1/\eta},
\ee
where $\alpha$ and $\beta$ are some constants.
Therefore,
\be                                                          \lb{eq17}
\sigma^{(ph)} = \frac{\alpha}{A^{\eta+1}} + 
\beta T^{1+1/\eta},
\ee
and this expression will reproduce as limiting cases  
both the already known theory of 
isentropic thin shells at $\beta=0$ and the thermodynamics of
a homogeneous 2D fluid 
$\sigma=\sigma(T)$ at $\alpha=0$.
Keeping in mind $\sigma^{(ph)}$, we nevertheless will attempt to 
construct the theory for arbitrary $f(A^\eta T)$.

First of all, from eq. (\ref{eq9}) one can obtain the relation between
entropy (which has to be a function of $A^\eta T$ as well) and $f$:
\be                                                          \lb{eq18}
A^\eta T \frac{\drm S}{\drm f}=1,
\ee
which (i) says that surface entropy can be found (up to additive constant)
if we know  $f$ explicitly, (ii) imposes, through the second law of
thermodynamics, the restriction of positive monotonicity of $f$:
$\drm f \geq 0$.
For example, following (i) for $f=f^{(ph)}$ we obtain 
\be                                                          \lb{eq19}
S^{(ph)} = \beta (\eta +1) A T^{1/\eta} + S_0,
\ee
where (ii) $\beta$ is restricted:
\be                                                          \lb{eq20}
\beta 
\left\{
       \begin{array}{l}
        \geq 0 ~\text{at}~ \eta \in (-\infty,-1) \cup [0, +\infty),\\
        \leq 0 ~\text{at}~ \eta \in (-1, 0),\\ 
        \text{arbitrary} ~\text{at}~ \eta = -1.
       \end{array}
\right.
\ee

Let us recall now the above-mentioned problem of the missing equation 
for the temperature field.
We will suppose the set of all the world sheet metrics (\ref{eq3}) and
accompanying fields to be a minisuperspace in the sense of the
Wheeler-DeWitt one.
It is evident that spherically symmetric world sheet metrics is 
determined by a single
function $R(\tau)$, and hence we can consider on this minisuperspace the
model described by the following action
\be                                                          \lb{eq21}
{\cal A} = \int {\cal L} \drm \tau,~~
{\cal L}=L w,
\ee
where $w=w(R,T)$ is an arbitrary gauge function, and
\be                                                        \label{eq22}
L = \frac{m \dot R^2}{2} 
- \frac{m}{2} 
\left[
       \Phi^- - 
       \left(
             \frac{\Delta\Phi - m^2/R^2}{2m/R}
       \right)^2
\right],
\ee
in which temperature and radius are 
considered as independent generalized coordinates.
Note that the gauge $w=1$ seems to be the most physically justified but
other gauges (e.g., $w=2 m_{\text{Planck}}/m$) are not evidently 
forbidden, and
therefore we will work with arbitrary $w$.

Extremalizing the action with respect to radius, 
$\delta {\cal A}/\delta R= 0$, we obtain the 
equation of radial motion
\be                                                        \label{eq23}
\frac{\drm (m w \dot R)}{\drm \tau}  = 
\frac{(m w)_{, R} \dot R^2}{2}
- w
\left[
       \frac{m \Phi^-}{2} - 
       \left(
             \frac{\Delta\Phi - m^2/R^2}{2\sqrt{2 m}/R}
       \right)^2
\right]_{,R},
\ee
where subscript ``$,x$'' means the partial 
derivative with respect to $x$.
Using time symmetry, we can decrease an order of this ordinary 
differential equation to obtain
\be                                                        \label{eq24}
\dot R^2  = 
\frac{2H}{m w} - \Phi^- +
       \left[
             \frac{\Delta\Phi - m^2/R^2}{2 m/R}
       \right]^2,
\ee
where $H$ is an integration constant.
Supposing it to be vanishing on real trajectories 
(thus one has the constraint $H\approx 0$) we obtain the equation of 
motion (\ref{eq6}).
Thus, our Lagrangian indeed describes dynamics of 
the layer
up to topological features which were described by 
the signs $\epsilon_\pm$.
However, we  can always restore the topology $\epsilon_\pm$
both at classical levels (rejecting redundant roots) and at quantum 
levels (considering appropriate boundary conditions for the 
corresponding Wheeler-DeWitt equation).

Further, extremalizing the action with respect to temperature, 
$\delta {\cal A}/\delta T= 0$, and taking into account
the equation of radial motion, we obtain the equation 
for the temperature field $T(R)$:
\be                                                        \label{eq25}
(\Phi^+ + \Phi^- - \phi) \phi_{,T} -
\left[
       \Delta\Phi^2 - 2 \phi (\Phi^+ + \Phi^-) + \phi^2
\right] (\ln{w})_{,T} = 0,
\ee
where $\phi=(4\pi\sigma R)^2=4\pi A^{-2\eta-1} f^2(A^\eta T)$.
It is clear that this expression is nothing but the constraint for
the temperature as a non-independent degree of freedom.
Therefore, for every physical concrete case
one should resolve eq. (\ref{eq25}) with respect to $T$ as a function of
radius and substitute it into the initial Lagrangian.
Thus, the radius remains the only canonical variable.
Further, 
one can see that gauge function $w$ does not affect the radial 
motion but appears in the temperature field equation.
(Moreover, it will be shown below that $w$ affects also on quantum
dynamics.)
Thus, the problem of obtaining a missing temperature equation
has been reduced to that of $w$ choice within the frameworks of the
minisuperspace model hypothesis.

Now we have all required expressions to perform the Wheeler-DeWitt
quantization of our model \cite{vil}.
Of course, such a quantization is not the only way
(see Ref. \cite{not}, and references therein).
However, in the absence of a rigorous axiomatic approach this method
has many advantages in comparison with others \cite{hkk}, as follows.
(i) Quantum dynamics can be constructed independently 
of time slicing on the basic patchwork manifold 
$\Sigma^+ \cup \Sigma \cup \Sigma^-$ (strictly speaking, such a 
union space does not have be the manifold in the conventional sense).
(ii) This method is simple and gives many heuristic results 
in a nonperturbative way, which is very important for non-linear
general relativity.
(iii) There explicitly exists conformity with the correspondence 
principle that improves physical interpretation of all the
concepts of the theory. 

According to eqs. (\ref{eq21}), (\ref{eq22}) the canonical
momentum conjugated to radius is 
\be                                                  \label{eq26}
\Pi_R = m w \dot R,
\ee
and hence the super-Hamiltonian has to be
\be                                                  \label{eq27}
{\cal H} = \Pi_R \dot R  - {\cal L} \equiv w H =0,
\ee
where the temperature-field constraint should be 
already taken into account.
In the quantum case $\hat\Pi_R = -i \partial/\partial R$ 
(we assume Planckian units) it yields 
the Wheeler-DeWitt equation for a wave function 
$\Psi (R,T(R))$ describing quantum oscillations of the layer surface,
\be                                                \label{eq28}
\frac{\partial^2 \Psi}{\partial R^2} + 
m^2 w^2
\left[
       \left(
             \frac{\Delta\Phi - m^2/R^2}{2m/R}
       \right)^2 -
       \Phi^- 
\right]\Psi = 0,
\ee
from which one can obtain eigenfunctions and spectra for all 
concrete physical values which can appear in the theory.

Let us summarize briefly the main points studied.
We generalized the theory of singular hypersurfaces for the case of
the nonconstant surface entropy and finite temperature.
Then we introduced the minisuperspace model which, firstly, provided
a variational procedure for describing the temperature field, and, 
secondly, determined all the canonical variables necessary for the
Wheeler-DeWitt's quantization of the theory.

\def\CJP{Czech. J. Phys.}
\def\CMPh{Commun. Math. Phys.}
\def\CQG {Class. Quantum Grav.}
\def\FP{Fortschr. Phys.}
\def\GRG {Gen. Relativ. Gravit.}
\def\JPh {J. Phys.}
\def\IJMP {Int. J. Mod. Phys.}
\def\LMPh {Lett. Math. Phys.}
\def\NPh  {Nucl. Phys.}
\def\PhE  {Phys. Essays}
\def\PhL  {Phys. Lett.}
\def\PhR  {Phys. Rev.}
\def\PhRL {Phys. Rev. Lett.}
\def\PhRp {Phys. Rep.}
\def\NCim {Nuovo Cimento}
\def\NuPB {Nucl. Phys.}
\def\prp {report}
\def\Prp {Report}

\def\jn#1#2#3#4#5{{(#5).} {\it #1} {\it #2}{\bf #3}, {#4}}

\def\boo#1#2#3#4#5{{(#4)}. {\it #1} ({#2}, {#3}){#5}}

\end{document}